\documentclass[amssymb, preprintnumbers, showpacs, showkeys, aps, prb, superscriptaddress, twocolumn, longbibliography]{revtex4-2}

\usepackage{color} 
\usepackage{xcolor} 
\usepackage[colorlinks,bookmarks=false,citecolor=blue,linkcolor=blue,urlcolor=blue]{hyperref}
\usepackage{slashed}
\usepackage{graphicx}
\usepackage{amsmath}
\usepackage{latexsym}
\usepackage{epstopdf}
\usepackage{amsmath}
\usepackage{enumitem}
\usepackage{amssymb,amsmath}
\usepackage{multirow}
\usepackage{mathtools}
\usepackage{bbm}
\usepackage{bm}
\usepackage{amsfonts}
\usepackage{amssymb}
\usepackage{calligra}
\usepackage{calrsfs}
\usepackage{makecell}
\usepackage{comment}
\usepackage[normalem]{ulem}
\usepackage[USenglish,american]{babel}
\usepackage{subfigure}

\expandafter\ifx\csname package@font\endcsname\relax\else
 \expandafter\expandafter
 \expandafter\usepackage
 \expandafter\expandafter
 \expandafter{\csname package@font\endcsname}%
\fi
\begin{document}

\title{Layering and superfluidity of soft-core bosons in shallow spherical traps}

\author{Fabio Cinti}
\email{fabio.cinti@unifi.it}
\affiliation{Dipartimento di Fisica e Astronomia, Universit\`a di Firenze, I-50019, Sesto Fiorentino (FI), Italy}
\affiliation{INFN, Sezione di Firenze, I-50019, Sesto Fiorentino (FI), Italy}
                
\author{Matteo Ciardi}
\email{matteo.ciardi@tuwien.ac.at}
\affiliation{Institute for Theoretical Physics, TU Wien, Wiedner Hauptstraße 8-10/136, 1040 Vienna, Austria}

\author{Santi Prestipino}
\email{sprestipino@unime.it}
\affiliation{Dipartimento di Scienze Matematiche e Informatiche, Scienze Fisiche e Scienze della Terra, Universit\`a degli Studi di Messina, viale F. Stagno d'Alcontres 31, I-98166, Messina, Italy}

\author{Giuseppe Pellicane}
\email{giuseppe.pellicane@unime.it}
\affiliation{Dipartimento di Scienze Biomediche, Odontoiatriche e delle Immagini Morfologiche e Funzionali, Universit\`a degli Studi di Messina, I-98125, Messina, Italy}
\affiliation{School of Chemistry and Physics, University of Kwazulu-Natal, 3209 Pietermaritzburg, South Africa}
\affiliation{National Institute of Theoretical and Computational Sciences (NIThECS), 3209 Pietermaritzburg, South Africa}

\begin{abstract}

Fundamental theories and models of many-body physics can be probed in experiments on ultracold atoms held in place by electromagnetic fields.
In particular, of considerable interest are systems under curved confinement, since they can yield exotic states of matter which would be impossible to obtain in flat space.
In this study we focus on relatively small samples, where curvature effects are stronger, and analyze by Monte Carlo simulations the peculiar structure arising in an assembly of soft-core bosons subject to a {\em weak} trapping potential with spherical symmetry.
Upon suitable tuning of the parameters, a hundred particles or so group together in clusters arranged in a shell with icosahedral symmetry. 
As the number of particles increases, a second shell gradually develops, concentric to (and partly overlapping with) the original one, where clusters are in perfect registry with the first shell, thus forming a dodecahedral pattern.
Cluster arrangements with the symmetry of other polyhedra are seen for different sets of parameters.
At low temperature the superfluid density is non-uniform in the radial direction;
heating the system progressively, superfluidity eventually vanishes while still clusters are present, a behavior resembling the transition from supersolid to normal solid on a plane.
Two shells of clusters are also observed in systems of classical or distinguishable quantum particles, but in those cases the shells are more fragile to thermal fluctuations.
All these behaviors can in principle be tested in systems of Rydberg-dressed atoms loaded into a bubble trap.
\end{abstract}

\date{\today}
\maketitle

{\em Introduction}---Among the main goals of quantum physics are the characterization and understanding of emergent behaviors in many-body systems.
On the experimental side, this objective is normally pursued by the use of ultracold-atom platforms~\cite{jaksch1998,greiner2002,Bloch2008}.
By now, thanks to the progress attained in confinement and cooling techniques, combined with the ability to tune particle interactions through Feshbach resonances, these devices (also referred to as ``quantum simulators'') are an ideal setup to test predictions of fundamental physics theories.
Prior and in parallel to experiments, numerical simulation~\cite{ceperley1995,krauth2006,gubernatis2016,becca2017} ensures a thorough analysis of the effects made by theoretically-informed interactions mimicking those of real systems, in this way allowing for a microscopic control of the ongoing phenomena.

An attractive confining geometry is the spherical surface, which offers the chance to explore, in diverse quantum systems, the relationship between curvature and correlation effects.
Even though a specific magnetic trap (the bubble trap~\cite{zobay2001,zobay2004,garraway2016,veyron2025}) has been invented for this purpose, a few technical difficulties --- above all, the need for microgravity conditions~\cite{aveline2020,carollo2022,lundblad2023} --- have so far prevented it from becoming a practical methodology.
However, this obstacle has not stopped theoretical investigation, which in the last few years has been primarily focused on the nature of the condensation and superfluid transitions in weakly-interacting bosons on a sphere~\cite{bereta2019,tononi2019,Tononi2024,tononi2020}.
In our recent work, we have rather concentrated on much denser samples of soft-core (i.e., penetrable) bosons~\cite{prestipino2019a,ciardi2024,PhysRevB.111.024512}, an important class of systems showing similarities to Rydberg atoms~\cite{henkel2010}, finding a rich low-temperature behavior with both uniform and non-uniform superfluid states.
In particular, the ``supersolid'' phases --- combining a spatially modulated density with superfluid response~\cite{pomeau1994,boninsegni2012a} --- are organized in clusters of overlapping particles, just as it happens on a plane~\cite{cinti2010,kunimi2012,macri2013,cinti2014a,prestipino2018}.
Recently, supersolidity has been investigated in various platforms, including dipolar condensates~\cite{chomaz2023,Casotti2024,PhysRevLett.128.195302,Tanzi2019,Biagioni2024,zhang2019,PhysRevA.104.013310,PhysRevLett.135.223402}, light-mediated interacting systems~\cite{Trypogeorgos2025Nature, atoms9030035}, as well as excitonic systems~\cite{PhysRevLett.132.176001}.

\begin{figure*}[t]
\begin{center}
\includegraphics[width=\linewidth]{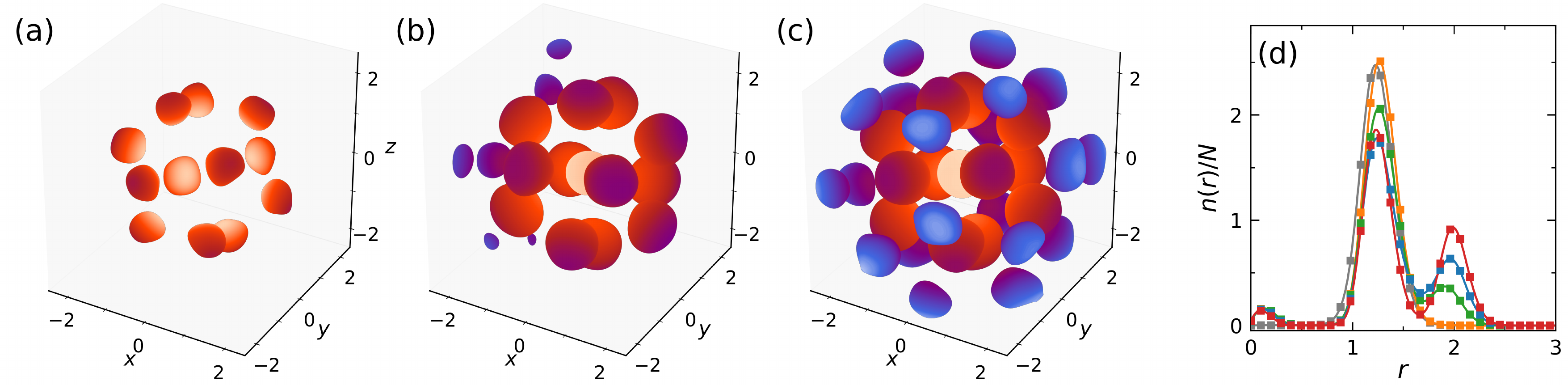}
\caption{Structure of bosons at $R=1.15,T=0.5$, and $\lambda=0.16$.
(a--c): Isodensity surfaces of the bosons at $N=200$ (a), $N=400$ (b), and $N=600$ (c).
Clusters are red in the first shell and blue in the second shell;
the pink cluster lies at the center of the trap.
(d): Integrated radial density $n(r)$ (see text) at discrete values $r$ of the distance from the trap center.
Lines are a guide for the eye:
$N=200$ (grey), 300 (orange), 400 (green), 500 (blue), and 600 (red).
To facilitate the comparison between different system sizes, each $n(r)$ has been divided by $N$.}
\label{fig1}
\end{center}
\end{figure*}

In a thin spherical shell the number and overall structure of clusters at equilibrium are determined by the sphere radius $R$ and, to a lesser extent, by the ``quantumness'' $\lambda=\hbar^2/2m$ ($m$ being the mass of particles), since the intercluster spacing is mostly dictated by the core diameter $\sigma$~\cite{ciardi2024}.
The interplay between $R/\sigma$ and $\lambda$ fix the polyhedral pattern of the cluster superstructure.

As long as the trapping potential is sharp around the minimum, clusters remain confined to a single shell.
On the other hand, we may wonder what happens near zero temperature ($T=0$) when the curvature of the potential at the minimum is sensibly reduced.
As the number $N$ of bosons increases, one possibility is that the population of clusters grows unlimited, while remaining confined near the surface of the reference sphere, at distance $R$ from the center.
On the other hand, a radially-dispersed system will pay the penalty of a larger trapping energy so as to reduce, to a greater extent, mutual interaction energy.

When the confining potential is weaker, particles are free to explore the radial direction to reduce interaction energy. While for small $N$ we still expect to find the same spherical shell of clusters, as $N$ increases we should see the emergence of truly three-dimensional structures. Due to the nature of the confining potential, which is harmonic around the minimum, we may expect clusters to break up and distribute on a thicker layer around $R$. Surprisingly, we instead find that the system favors distributing the new particles on a second spherical shell at larger radius, while leaving the first shell largely unchanged, despite the potential growing as $r^2$.

{\em Results}---In this letter, we investigate a system of interacting spinless bosons described by the following Hamiltonian:
\begin{equation}
H=\sum_{i=1}^N\left(-\lambda\nabla^2_i+u_{\rm ext}(|{\bf r}_i|)\right)+\sum_{i<j}u_{\rm int}(|{\bf r}_i-{\bf r}_j|)\,,
\label{eq1}
\end{equation}
where the two-body interaction potential $u_{\mathrm{int}}(r)$ is modeled as a square barrier of height $\epsilon$ and width $\sigma$, while
\begin{equation}
u_{\rm ext}(r)=(u_0/\Omega)\sqrt{(r^2-R^2)^2/4+\Omega^2}-u_0
\label{eq2}
\end{equation}
represents the bubble-trap potential.
In Eq.~(\ref{eq2}), $u_0$ and $\Omega$ are parameters characterizing the trap~\cite{sun2018}.
In the following, we take the parameters of the soft-core potential, $\epsilon$ and $\sigma$, as our energy and length units.
Temperatures are also in units of $\epsilon$, with $k_B=1$.
Leveraging on the isomorphism between bosons and imaginary-time paths (worldlines)~\cite{feynman2010}, our simulations are made using the path integral Monte Carlo (PIMC) method~\cite{ceperley1995}, along with the worm algorithm~\cite{boninsegni2006b} for a more efficient sampling of particle permutations.
For our demonstration, we take the radius of the reference sphere to be $R=1.15$, so as to stabilize, for $u_0=1$ (i.e., a low value), $\Omega=0.0441,\lambda=0.16,T=0.5$, and $N=200$, the icosahedral supersolid with 12 clusters.
Then, we increase $N$ in steps of 50, keeping all the other parameters fixed, to see how the equilibrium structure changes.
In the presence of clusters, the number density $\rho$ depends on $r$, $\theta$ and $\phi$. For clarity, when showing density profiles as a function of the radial coordinate, we present the integrated radial density $n(r) = r^2 \int{\rm d}\theta\,{\rm d}\phi \, \sin \theta \, \rho(r, \theta, \phi)$, which corresponds to the number of particles in an infinitesimal spherical shell at $r$ (with $N = \int{\rm d}r \, n(r)$).

In Fig.~\ref{fig1}(a--c) we draw the three-dimensional density at $N=200$, 400, and 600, showing the transition from a shell with 12 clusters at $N=200$ to the formation of a second shell at $N=600$. To better understand the formation of the second layer, in Fig.~\ref{fig1}(d) we plot $n(r)/N$ at different $N$. We observe the onset of a second shell between $N=300$ and 350. The transition is not abrupt, and clusters in the second shell are filled up gradually as we increase the number of particles.

\begin{figure}[t!]
\includegraphics[width=9cm]{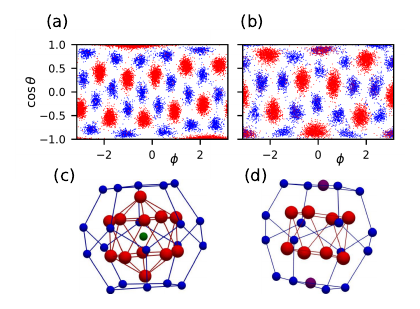}
\caption{(a): Typical equilibrium configuration for $N=600, R=1.15,T=0.5$, and $\lambda=0.16$ ($\theta$ is the polar angle and $\phi$ is the azimuthal angle).
Red (blue) symbols correspond to beads in the first (second) shell, while the beads in the central cluster have been omitted for clarity.
(b): The same as (a), but for $N=480$, $R=1.05,T=0.7$, and $\lambda=0.16$.
(c--d): Simplified 3D representation of the cluster structure in (a) and (b), respectively.
Clusters are generated through a thresholding algorithm and their size is suggestive of the number of particles within.
Lines are a guide for the eye, pointing out the formation of different polyhedral structures.}
\label{fig2}
\end{figure}

For $N=600$ the configuration of worldline beads, like the one reported in Fig.~\ref{fig2}a, reveals the presence of 20 clusters in the outer shell, which, by visual inspection (Fig.~\ref{fig2}c), are located at the vertices of a dodecahedron (the dual polyhedron of the icosahedron).
The concurrence of icosahedral and dodecahedral order in a single structure is a gratifying example of complex architecture emerging from scratch, based uniquely on the principle of ground-state energy minimization.
The average cluster population is different in the two shells:
for $N=600$, it is about three times larger for the clusters in the first shell than for those in the second shell.
Moreover, another small cluster appears at the center of the trap, facilitated by the finite height of the confining potential at $r=0$.
If $N$ were to grow much beyond 600 we expect the eventual formation of a third shell of clusters.

The icosahedron-dodecahedron combination is maximally symmetric and, for this reason, robust. In the Supplemental Material \cite{supplemental} we provide a consistent rationalization of the evolution of system structure with $N$ in terms of a classical lattice-gas model, using the minimization of grand potential at $T=0$ as the only guiding principle, which further confirms the structural stability of our quantum samples.

However, other structures will emerge for different sets of parameters.
For example, in the right panels of Fig.~\ref{fig2} we draw the typical equilibrium configuration of a system with 480 bosons, for $R=1.05$.
The smaller radius brings the system to accommodate fewer clusters in the first shell, but again arranged in a symmetric fashion.
Notably, the clusters sit at the vertices of a capped square antiprism --- though the ``caps'' are actually located in the valley between the shells, and thus shared between them.

Two-shell ordering is not exclusive to bosons, suggesting a geometric origin for the observed symmetry.
Indeed, by developing an ad hoc classical benchmark~\cite{supplemental} we recover the same icosahedron-dodecahedron structure of quantum clusters in molecular-dynamics simulations of a classical system of 600 penetrable spheres with $R=1$. Crucially, for a fixed core diameter $\sigma$, a smaller radius is required in the classical case to achieve the same arrangement: this highlights how quantum delocalization effectively increases the particle ``size" relative to the nominal diameter.
Though identical to bosons, however, the cluster structure of the classical system --- and of boltzmannons as well~\cite{supplemental} --- is more fragile, as it melts at a lower temperature. 
This feature is also present in another model of penetrable bosons~\cite{cinti2014b}. 

\begin{figure}[t]
\centering
\includegraphics[width=0.51\textwidth]{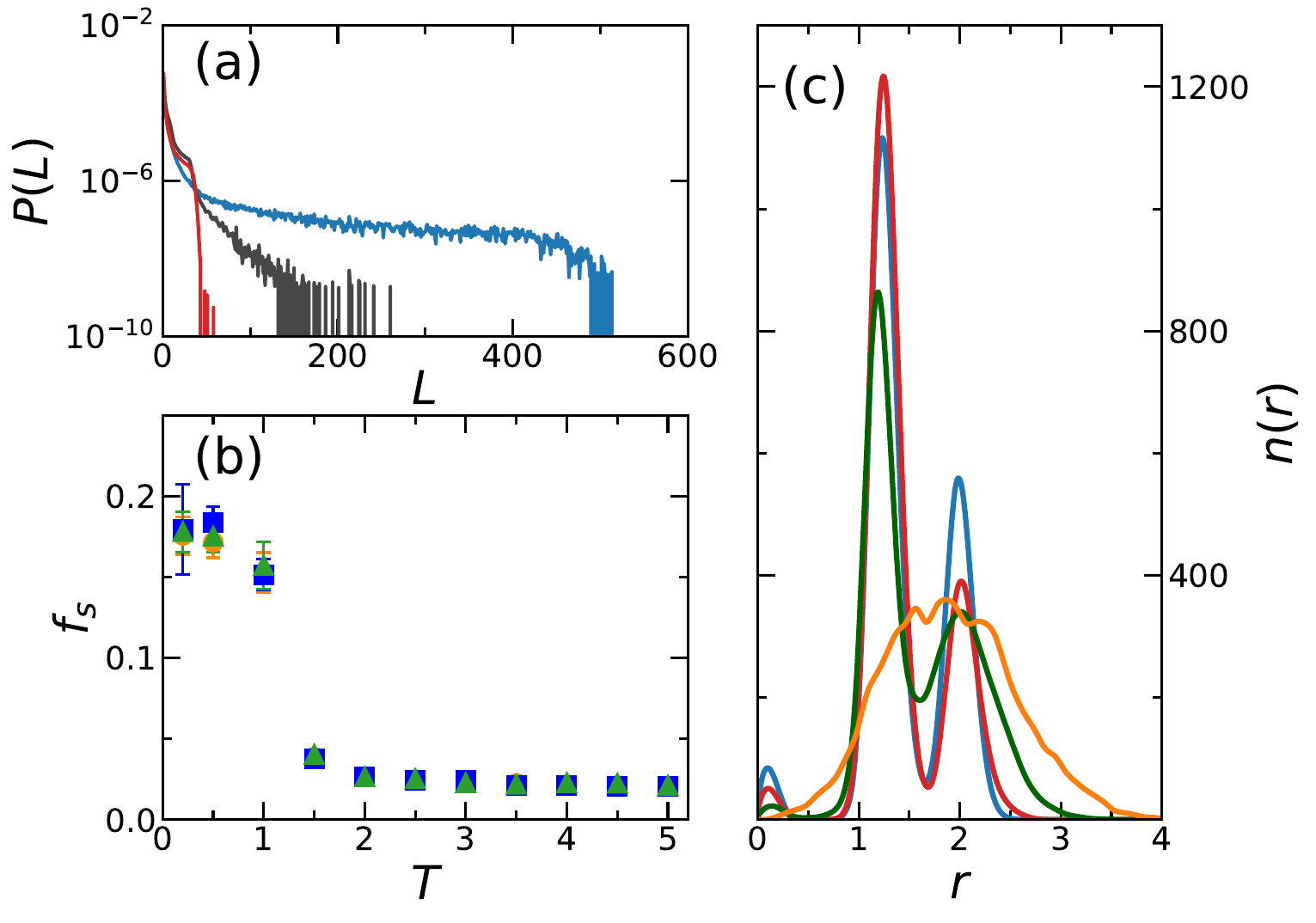}
\caption{System with $N=600$ bosons at $R=1.15$ and $\lambda=0.16$.
(a): Distribution $P(L)$ of cycle lengths (see text) for $T=0.5$ (blue), 1.5 (grey), and 5 (red).
(b): Components of the superfluid fraction $f_s$ along three mutually-orthogonal directions.
(c): Integrated radial density $n(r)$ for $T=0.5$ (blue), 5 (red), 10 (green) and 20 (orange).}
\label{fig3}
\end{figure}

Along with the structure, we have also analyzed the statistics of the length $L$ of exchange cycles $P(L)$, defined as the probability that a particle chosen at random belongs to a cycle of length $L$ \cite{jain2011a}.
A long-tailed distribution is indicative of global coherence throughout the system, implying a non-zero condensed fraction and superfluidity.
In this case, the longest cycles will embrace beads from clusters in both shells.
In Fig.~\ref{fig3}a we compare the distribution of cycle lengths at different temperatures, for $N=600$.
At the lowest $T$, lengths from 400 to 500 have a non-zero occurrence, giving support to the idea that the system is superfluid.
Upon heating, cycles become progressively shorter, suggesting a gradual extinguishing of superfluidity which remains confined to individual droplets.

The superfluid fraction, $f_s$, is related to the projected area enclosed by imaginary-time paths~\cite{sindzingre1989}, a quantity that can be evaluated in PIMC simulations.
As a response function, $f_s$ can be defined along three orthogonal directions; we plot all components in every plot, finding perfect agreement between them in every case.
Our estimate of $f_s$ is plotted as a function of $T$ in Fig.~\ref{fig3}b. We have checked that, at the lowest $T$ used, $f_s$ is independent of the number of time slices employed in the simulation.
Beside confirming that at $T=0.5$ the system is indeed superfluid ($f_s\simeq0.175$), hence supersolid, we also see that superfluidity vanishes at $T\simeq 1.5$.
The behavior of $f_s(T)$ should be contrasted with the greater robustness of two-shell structure, which survives for temperatures much higher than $T=1$ (see Fig.~\ref{fig3}c).

\begin{figure}[t]
\centering
\includegraphics[width=\linewidth]{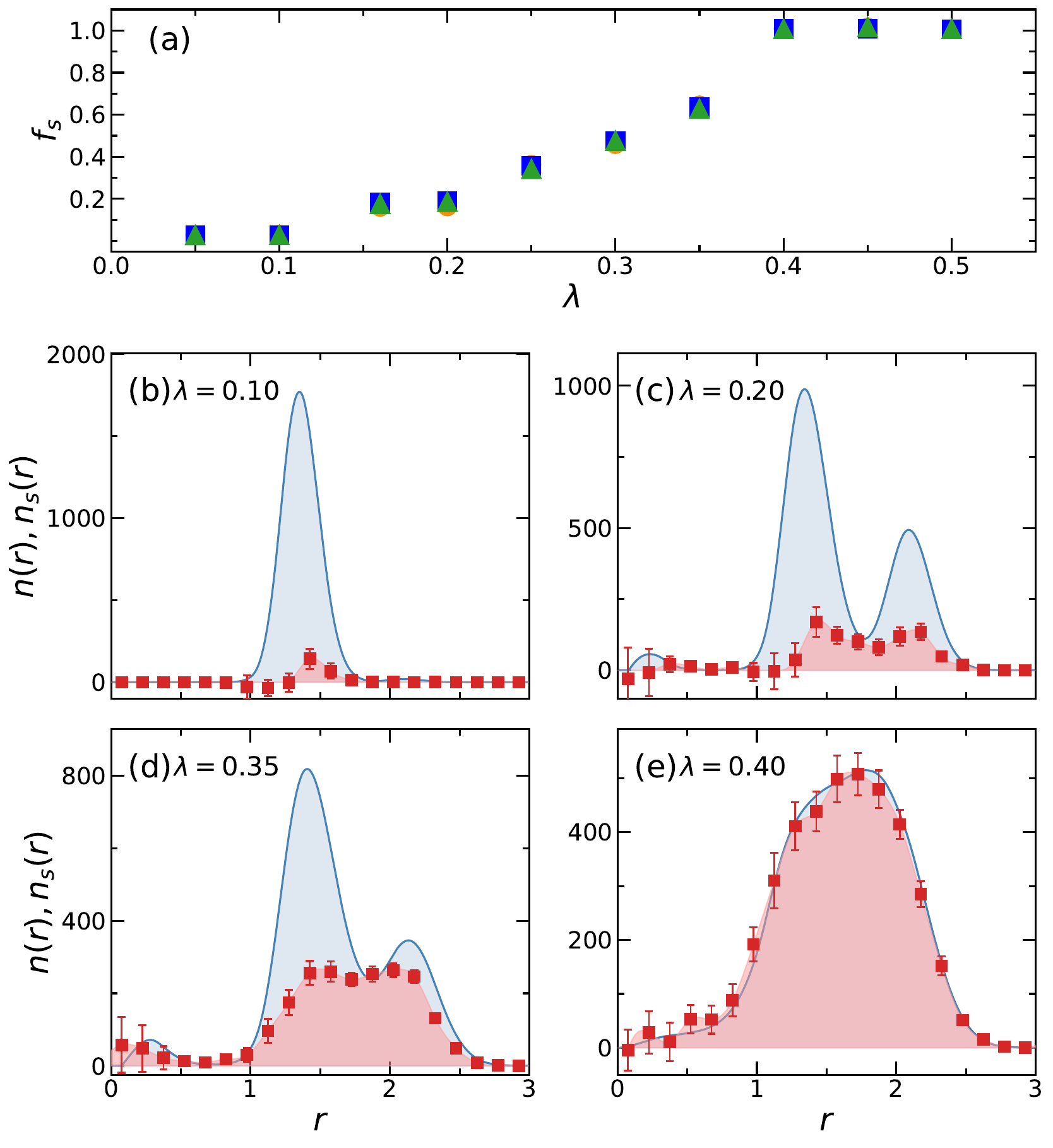}
\caption{System with $N=600$ bosons at $R=1.15$ and $T=0.5$.
(a): Components of the superfluid fraction $f_s$.
(b--e): Integrated superfluid radial density $n_s(r)$ (see text) vs. integrated radial density $n(r)$ at various $\lambda$.}
\label{fig4}
\end{figure}

It is actually possible to define a local measure of superfluidity~\cite{ciardi2022b}, obtained by histogramming the radial dependence of path area, to see where superfluidity is mostly concentrated in the system (see more in Ref.~\onlinecite{supplemental}).
We have thus divided space in thin spherical slices and computed the contribution to $f_s$ from each of them.
This allows us to compute the integrated superfluid radial density $n_s(r)$ and compare it with $n(r)$.
In Fig.~\ref{fig4} we look at the dependence of structure and superfluidity on the mass of bosons, i.e., on the magnitude of quantum fluctuations, for $N=600$ and $T=0.5$. Additional plots of the cluster structure are found in the Supplemental Material \cite{supplemental}.
We identify three distinct regimes:
For $\lambda$ smaller than $\approx 0.1$ there is only one shell (Fig.~\ref{fig4}b) and the behavior is essentially the same of distinguishable quantum particles ($f_s\simeq 0$).
By increasing quantum fluctuations up to $\lambda\approx 0.35$, the shells become two (Fig.~\ref{fig4}c and d) and $f_s$ grows with $\lambda$ until about $0.6$.
In this region the system is supersolid.
The overall profile of $n_s(r)$ is similar to $n(r)$, but in the central cluster and in the valley between the shells superfluidity is, on a percentage basis, higher. It is interesting to note that, as quantum fluctuations grow, the superfluid density between the shells increases accordingly.
For still larger $\lambda$, clusters are no longer present~\cite{supplemental} and particles are now dispersed in a broad superfluid shell ($f_s\simeq 1$), see Fig.~\ref{fig4}.

{\em Discussion}---
We find out that a few spherical layers of atomic clusters can be held in place by a weak bubble-trap potential --- i.e., by an external force that pulls particles gently toward $r=R$.
This outcome bears similarities to the phenomenon of atomic layer deposition, where a thin film is grown on a crystalline surface by exposing gaseous atoms to an attractive substrate.
In our setting, a low-temperature gas of $N=200$ particles forms a thin shell at $r\simeq R$.
By progressively increasing $N$ up to 600, we observe the building up of a second shell of larger radius.
Additional shells are expected to form when the chemical potential of the gas grows further.

Our simulation employs spinless bosons with a soft core, so as to induce a double level of spatial organization in the system:
In worldline representation, particle beads first gather together in disordered clusters, and clusters (which behave like bigger atoms) then self-assemble in two curved ``crystals'', nested in each other, made by the vertices of a convex polyhedron.
Among the possibilities, we identify two notable cases:
A commensurate situation where the two polyhedra (icosahedron and dodecahedron) are perfectly interlocked, since dual to each other, and a less symmetric case where the inner polyhedron is a semiregular capped square antiprism.

The added value of our study is the quantum character of system particles, which brings about remarkable implications for the nature of the system ``phases''.
By measuring the superfluid fraction we establish that, at the lowest temperatures probed, the bilayer is indeed supersolid.
In this case, the longest imaginary-time paths involve beads from both shells, implying extended quantum coherence throughout the system. 
Upon heating, superfluidity eventually vanishes, and the loss of superfluid response vastly precedes the evaporation of first-shell clusters.
While thermal fluctuations eventually lead to the destruction of solid order, quantum fluctuations have the opposite effect, pushing particles to the outer shell and leading to the formation of a second structured layer.
The behaviors just described can in principle be tested in systems of Rydberg-dressed atoms loaded into a bubble trap.
Moreover, the dynamic signatures of the present supersolid can be unveiled in traps equipped with rotating potential barriers~\cite{nesti2026}.

The structure of nested polyhedra in the bubble trap is reminiscent of that observed in so-called {\em snowballs} forming around ionic impurities in bulk superfluid helium~\cite{Buzzacchi2001, Galli2011, Tramonto2015}. Successive shells in helium snowballs form gradually as more He atoms are added around the ion, with little back-action from the external shells onto the inner ones.
Here, despite the differences in both external potential and interaction, we observe a similar behavior, with the first shell remaining centered around $R$, while the second shell builds up at a larger radius.
A clear difference is that, in the present case, each vertex of the polyhedron is occupied by a particle cluster, rather than the single atoms forming the inner shells of helium snowballs.
The clusters we observe are also quite different from the ones forming in the outer shells of helium snowballs, which occur as a density modulation on top of a superfluid background, approaching the bulk limit.
The internal structure of helium snowballs is also entirely determined by the fixed He-He and ion-He interaction, whereas here it can be manipulated by altering the trap, allowing the exploration of a range of different geometries within the same setup.

\vspace{2mm}
{\em Acknowledgments}---This work was supported by the European Union through the Next Generation EU funds through the Italian MUR National Recovery and Resilience Plan, Mission 4 Component 2 - Investment 1.4 - National Center for HPC, Big Data and Quantum Computing (CUP B83C22002830001).
F.~C. acknowledges financial support from PNRR MUR Project No. PE0000023-NQSTI.
M.~C. acknowledges funding from the Austrian Science Fund (Grant No. 10.55776/COE1).
F.~C. and G.~P. acknowledge the NICIS Centre for High Perfomance Computing, South Africa, for providing computational resources.

\vspace{2mm}
{\em Data availability}---The data that support the findings of this article are available from the authors upon reasonable request.


%


\clearpage
\onecolumngrid

\begin{center}
    \textbf{\large Supplemental Material}
\end{center}

\setcounter{equation}{0}
\setcounter{figure}{0}
\setcounter{table}{0}
\setcounter{section}{0}
\setcounter{page}{1}

\renewcommand{\theequation}{SM\arabic{equation}}
\renewcommand{\thefigure}{S\arabic{figure}}
\renewcommand{\thesection}{\Roman{section}}



This document contains further numerical and analytical results on the boson system investigated in the main text. Section I provides some additional Path Integral Monte Carlo (PIMC) simulation data. The classical limit of the system is studied in Section II. In Section III we provide a simple theory of the dependence of system structure on the number of particles; the theory illustrates how, on increasing the chemical potential, we observe the sequential filling of two spherical layers having the symmetry of an icosahedron and a dodecahedron, respectively. Finally in Section IV, we derive expressions for the zonal superfluid fraction and the superfluid density.

\section{Additional PIMC results}

\vspace{5mm}
\begin{center}
{\bf I. Additional PIMC results}
\end{center}

\begin{figure}[b]
\includegraphics[width=0.8\linewidth]{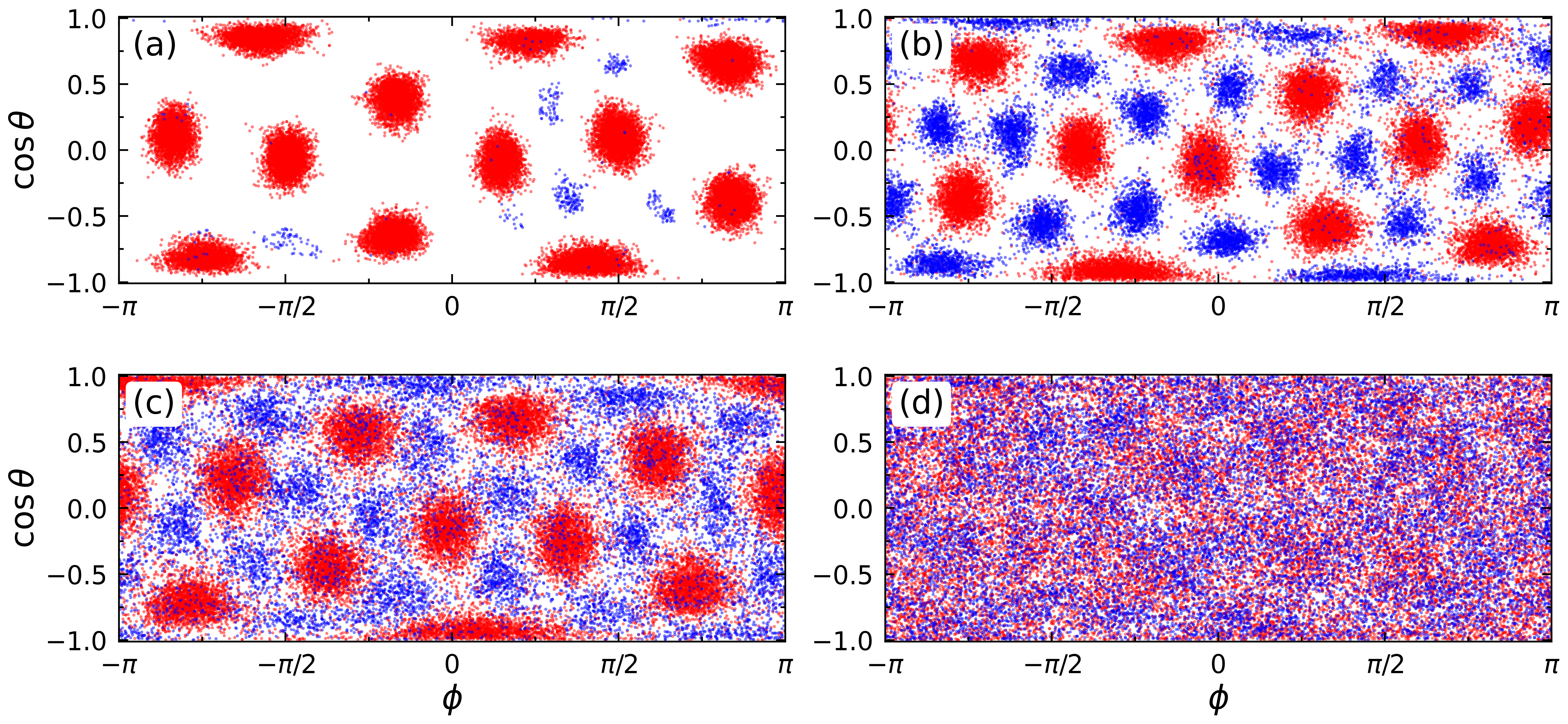}
\caption{Snapshots of worldline configurations projected onto the $\phi$-$\cos \theta$ plane, at four values of $\lambda$: 0.1(a), 0.2(b), 0.35(c), 0.4(d).}
\label{fig5}
\end{figure}

\begin{figure}[t]
\includegraphics[width=0.8\linewidth]{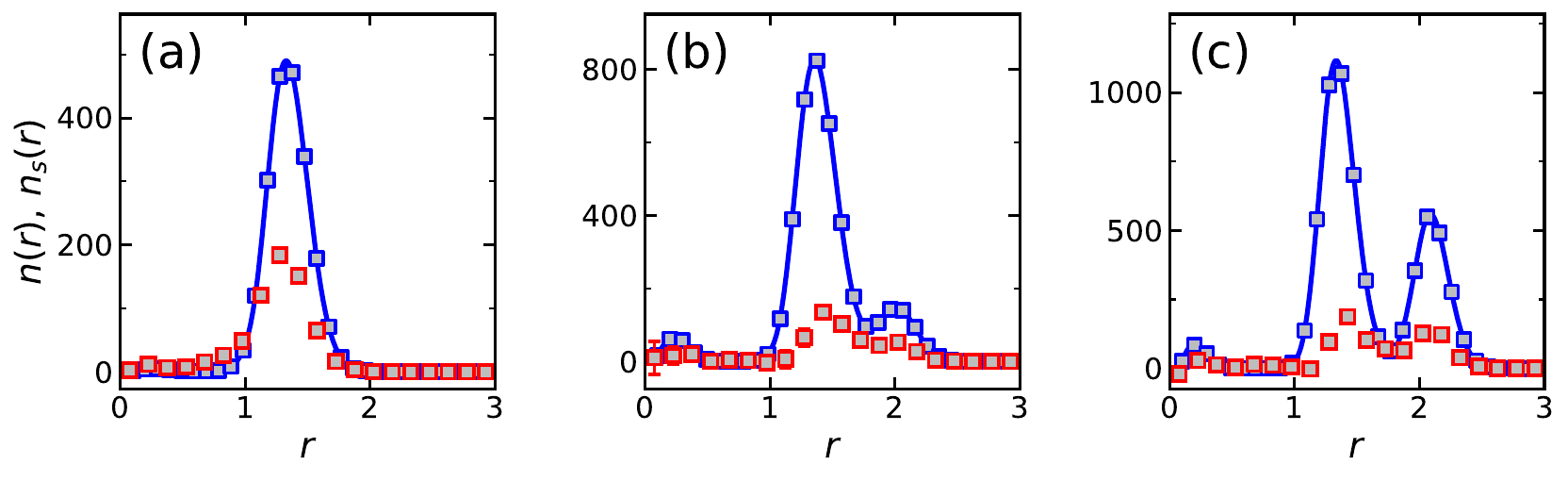}
\caption{Total (blue) and superfluid (red) integrated radial densities, at $R=1.15$, $T=0.5$, $\lambda=0.16$ and for three values of $N$: 200 (a), 400 (b), and 600 (c).}
\label{fig:nr_ns}
\end{figure}

\begin{figure}[t]
\includegraphics[width=0.5\linewidth]{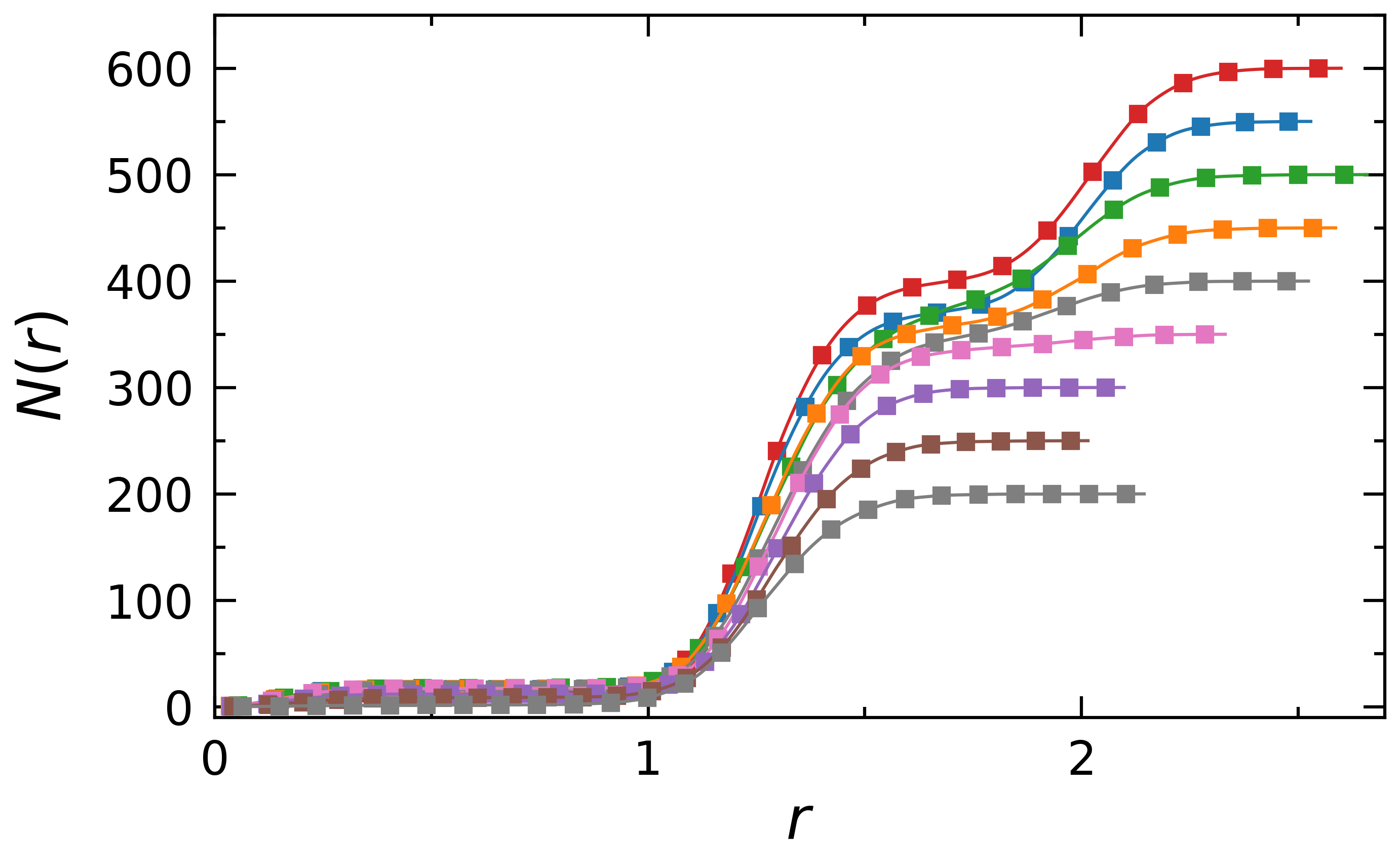}
\caption{Cumulative particle number as a function of $R$ for the values of $N$ in Fig.~\ref{fig1}, from $N=200$ to $N=600$ in steps of 50 particles. }
\label{fig:cumulative_nr}
\end{figure}

\begin{figure}[t]
\includegraphics[width=0.5\linewidth]{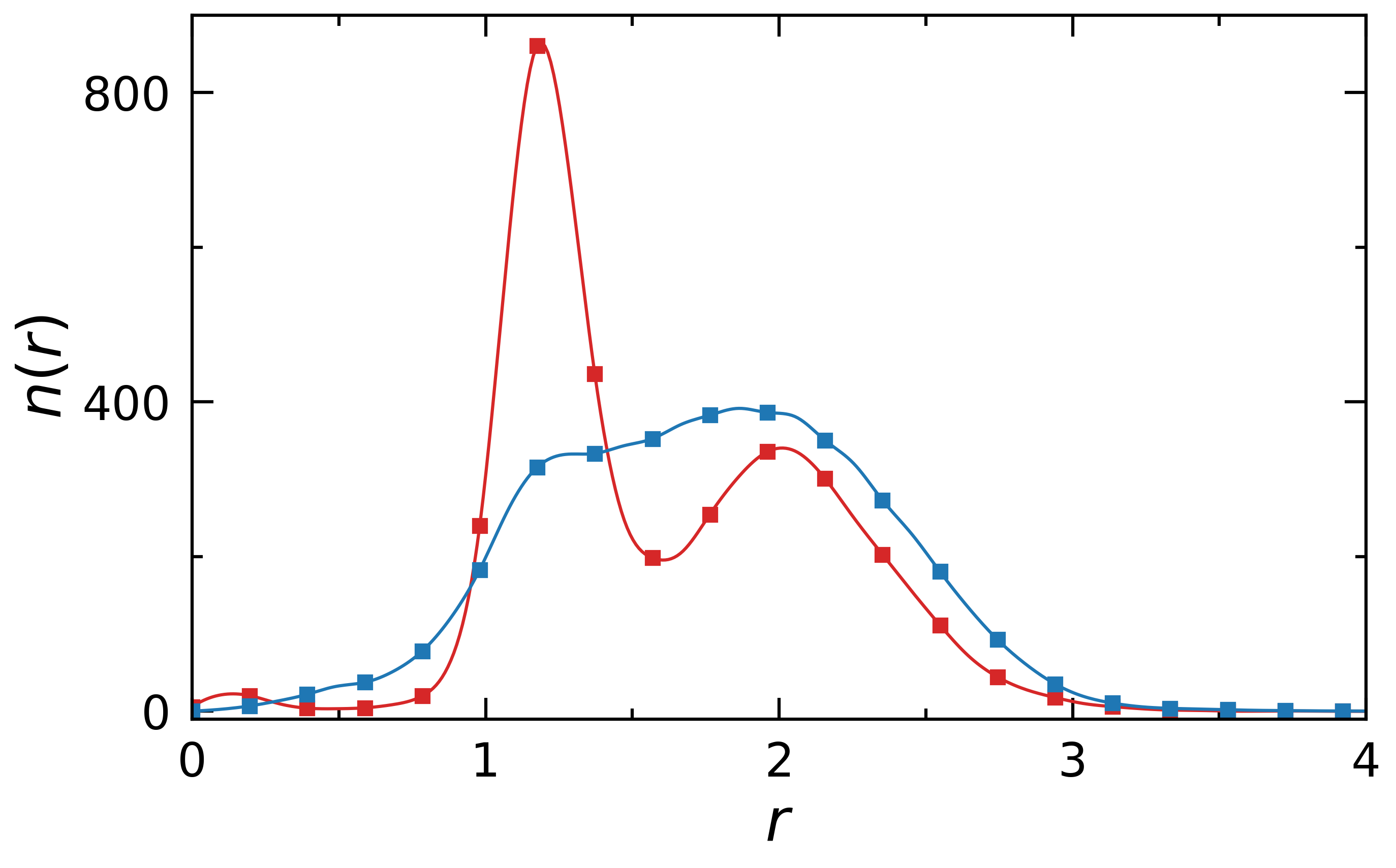}
\caption{Integrated radial density for $N=600$, $R=1.15$, and $T=10$, for bosons ($\lambda=0.16$) (red) and for boltzmannons with the same parameters (blue).}
\label{fig:bose_boltz}
\end{figure}

\begin{figure}[t]
\includegraphics[width=0.8\linewidth]{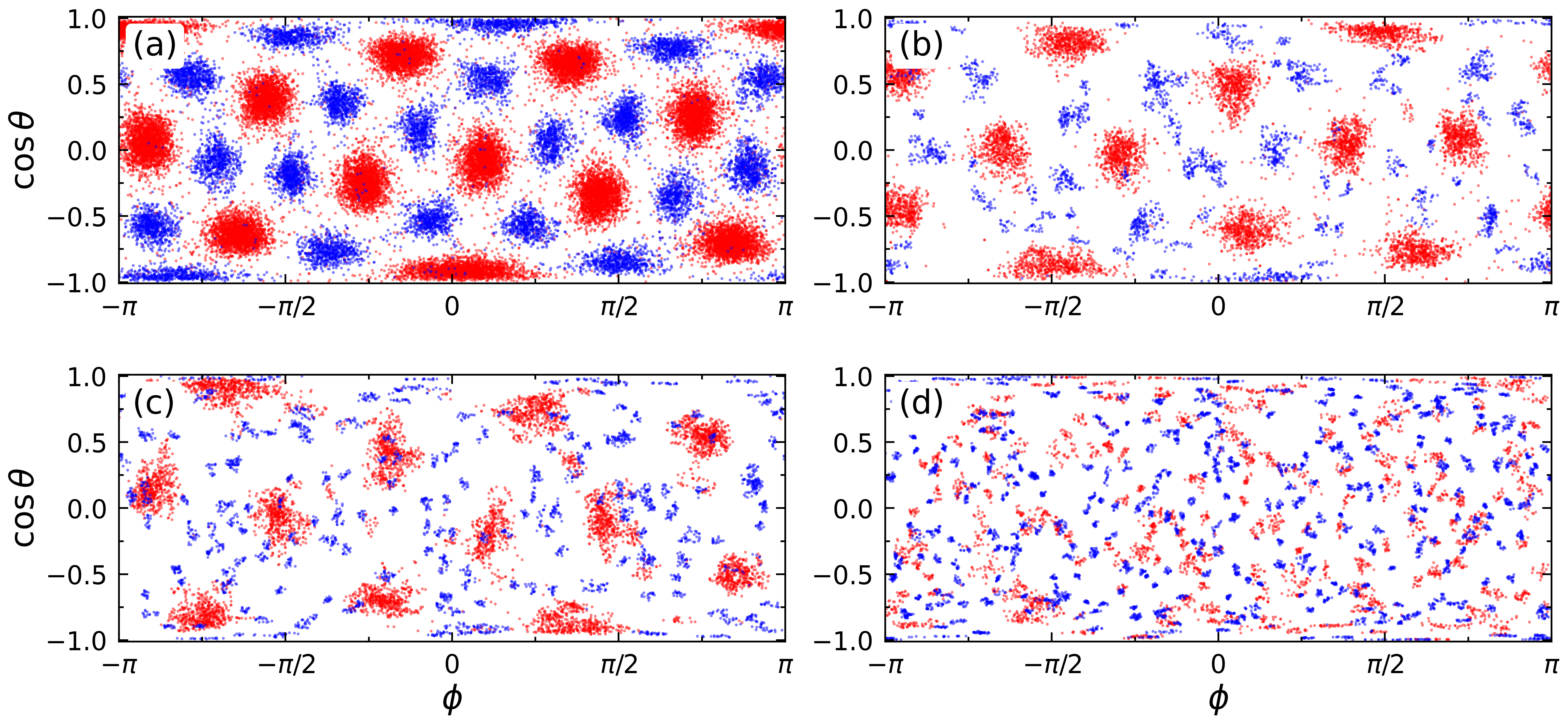}
\caption{Snapshots of typical worldline configurations projected onto the $\phi$-$\cos \theta$ plane for the same cases reported in Fig.~\ref{fig3}c. (a) $T=0.5$, (b) $T=5.0$, (c) $T=10.0$, (d) $T=20.0$.}
\label{fig:config3c}
\end{figure}

In Fig.~\ref{fig5}, we show worldline snapshots for the sets of parameters shown in Fig.~\ref{fig4}. The projections in the $\phi$-$\cos\theta$ plane reflect the transition seen in the integrated radial density profiles. More classical particles ($\lambda = 0.1$) form a single layer with twelve clusters; as the quantumness grows, clusters  only accommodate part of the particles and the rest are pushed out to the second layer. Clusters grow larger and more fluid, until the entire structure melts at $\lambda \approx 0.4$, seen in Fig.~\ref{fig4}(d).

In Fig.~\ref{fig:nr_ns}, we show the integrated radial density and integrated superfluid radial density for changing particle number. For $N=400$ and $600$, we note that the system results completely superfluid at the interface of the two shells, $n(r)\approx n_s(r)$.
This is clear evidence that the two shells are phase coherent between themselves.

In Fig.~\ref{fig1} of the main text, we show the integrated radial density profiles at various values of $N$, normalized so to show the gradual growth of the second shell. Here, in Fig.~\ref{fig:cumulative_nr} we present the same data from a different point of view, displaying the cumulative particle number $N(r) = \int_0^r dr' \, n(r')$ (which converges to $N$ once $r$ is large enough to include all particles in the system). This figure shows more clearly how the number of particles in the first shell continues to grow, even as the second shell is forming, and allows for a concrete measure of the number of particles in each shell, given by the particle numbers at which $N(r)$ flattens out.

In Fig.~\ref{fig:bose_boltz} we plot the histogram of particle number as a function of $r$ for bosons and for boltzmannons at $T=10$. We clearly see how bosons still maintain the double-shell structure (as already shown in Fig.~\ref{fig3}c of the main text), while, for boltzmannons, both the double-shell structure and the clusters are absent. We see therefore that Bose-Einstein statistics plays a significant role in stabilizing the geometry of the system \cite{cinti2014b}.

Finally, we provide in Fig.~\ref{fig:config3c} additional information on the thermal evolution of the boson system with $N=600,R=1.15$, and $\lambda=0.16$, showing snapshots of the system for the same set of temperatures as in Fig.~\ref{fig3}c. We clearly see that clusters are still present in the first shell for $T=10$, whereas they have already melted for $T=20$.


\vspace{5mm}
\begin{center}
{\bf II. Classical simulations}
\end{center}

\begin{figure}[t]
\includegraphics[width=5.cm]{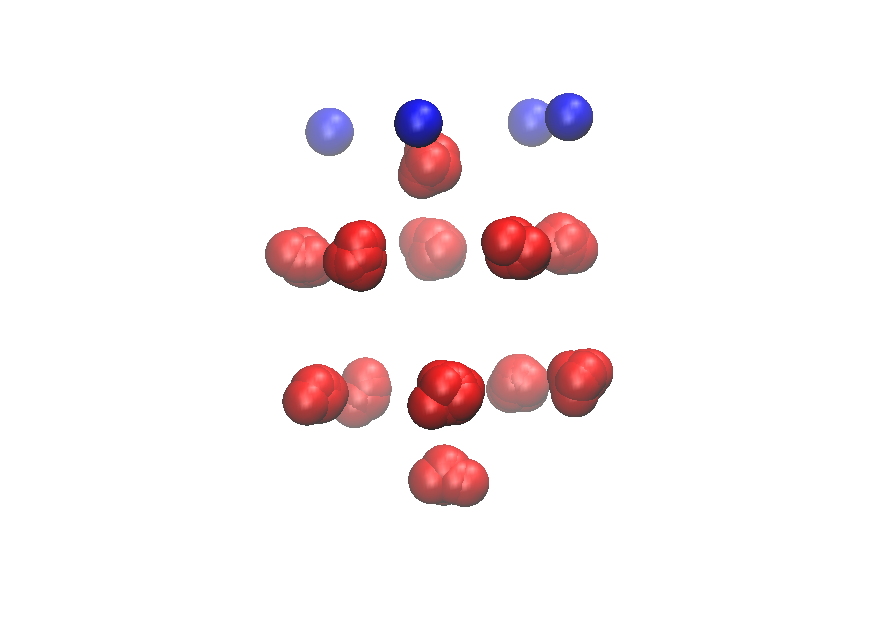}\qquad\qquad
\includegraphics[width=5.cm]{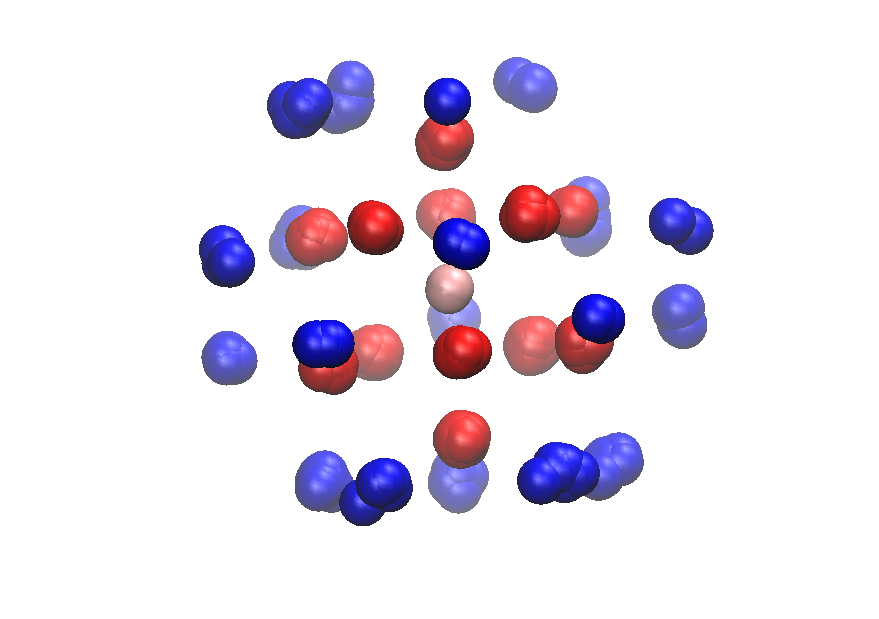}\qquad\qquad
\includegraphics[width=5.cm]{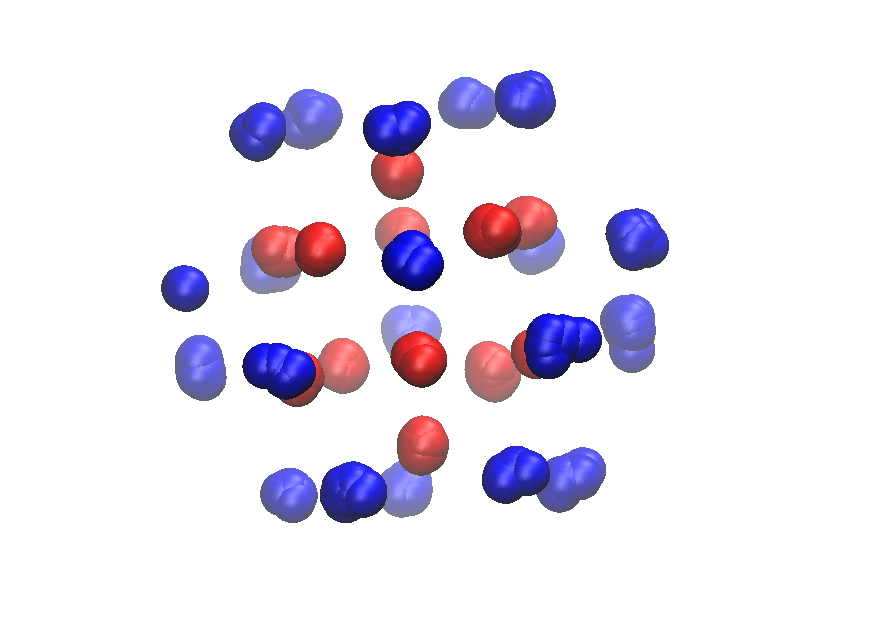}
\caption{Snapshots of typical configurations of the classical particles inside the simulation box (the symbol size is not in scale with the box length). Left: $N=200$; middle: $N=400$; right: $N=600$.}
\label{fig:classical_snaps}
\end{figure}

\begin{figure}
\includegraphics[width=0.7\linewidth]{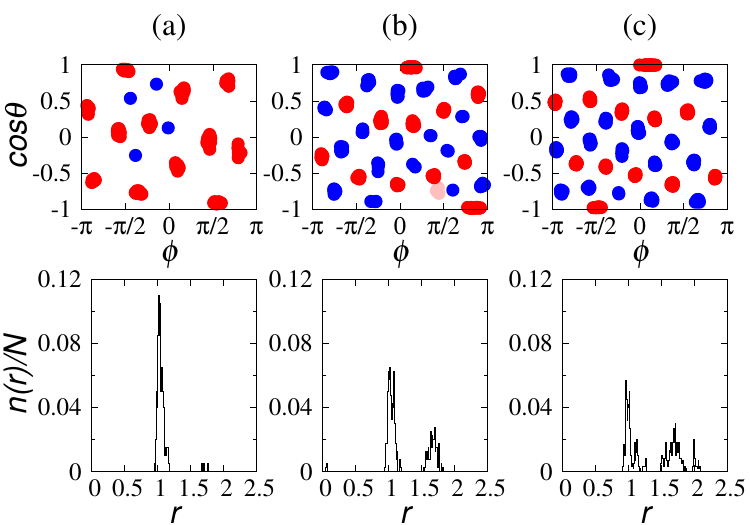}
\caption{Top panels: Snapshots of particle configurations projected onto the $\phi$-$\cos \theta$ plane. The pink blob corresponds to the cluster (made of four particles) lying near the center of the trap. Bottom panels: normalized histograms of the number of particles as a function of the distance $r$. (a) $N=200$ (b) $N=400$, (c) $N=600$. }
\label{fig:classical_histos_and_maps}
\end{figure}

In this section, we report on the results of molecular dynamics (MD) simulations of  penetrable spheres of diameter $\sigma$, subject to the external potential in Eq.~(\ref{eq2}). For the sake of using a simple algorithm for the numerical integration of the equations of motion (and avoiding to deal with impulsive forces, i.e., sharp collisions), we actually employ a continuous generalized exponential model potential with exponent $30$ and height $\epsilon$, whose harsh steepness in the neighborhood of $\sigma$ makes it almost indistinguishable from a square barrier. Simulations are performed at fixed $N=200, 400, 600$ and $T=0.5$, for $u_0=1$ and $\Omega=0.0441$, while the reference radius $R$ is smaller than the one adopted in quantum simulations, i.e., $R=1$. The timestep is $dt=0.005\sqrt{\frac{m\sigma^2}{\epsilon}}$ and the thermal bath is modeled by a Nos\'e-Hoover thermostat with a temperature damping factor $\tau=100 dt$ in time units. Regarding the initial configuration, particles are placed at random on the reference sphere of radius $R$. We let the system evolve for tens of million MD steps, though the equilibrium configuration is actually reached in the first thousands of MD steps. Initial velocities are randomly selected from a Gaussian distribution. 

Regardless of the value of $N$, the system eventually reaches a clustered configuration, and the phenomenology of thermalization closely resembles the results reported in Fig.~\ref{fig1}(a-c). At $N=200$, clusters are organized into a regular icosahedral configuration at distance $r\approx 1$ from the center of the sphere, as shown in the left panel of Fig.~\ref{fig:classical_snaps}, showing the typical equilibrated configuration in the simulation box. We provide additional evidence of twelve-clusters' ordering on a spherical surface in the left panels of Fig.~\ref{fig:classical_histos_and_maps}, where we show the $\cos\theta-\phi$ map of particle positions and the normalized histogram of the number of particles ($n(r)/N$). The icosahedral ordering in the first shell keeps unchanged with increasing $N$; however, when $N$ doubles to 400, on top of the first shell a second layer of clusters develops at $r \approx 1.7$, as shown in the middle panels of Fig.~\ref{fig:classical_snaps} and \ref{fig:classical_histos_and_maps}. The second layer of clusters shows dodecahedral order, even though not complete because the number of clusters is 19 rather than 20 (see the top middle panel of Fig.~\ref{fig:classical_histos_and_maps}). Finally, at $N=600$ dodecahedral order is perfect, as can be seen in the right panels of Fig.~\ref{fig:classical_snaps} and \ref{fig:classical_histos_and_maps}.

Although classical simulations confirm the picture of a progressive building up of the second layer with increasing $N$, same as reported in quantum MC simulations, there are some quantitative differences which can be traced back to the enhanced steric effect of quantum correlations. Indeed, to observe the same phenomenology of bosons we had to reduce the surface area by nearly a $25\%$, setting $R=1$. We also note that the population of clusters in the second shell increases faster with $N$ in the classical simulations, but this effect should possibly be ascribed to the choice of a different $R$.
The onset of the second shell is already visible for $N=200$, as can be seen in the left panels of Fig.~\ref{fig:classical_histos_and_maps}. However, the full dodecahedral structure is only appearing at $N=600$ (note the close similarity of quantum and classical configurations, as reported in Fig.~\ref{fig1}c and in the right panel of Fig.~\ref{fig:classical_snaps}).
Finally, at $N=600$ MD simulations show hints of a third shell at a germinal stage, with only a dozen of particles, as reported in the histogram of the bottom panel of Fig.~\ref{fig:classical_histos_and_maps} (we decided not to report the few third-shell particles in the snapshot of the simulation box and in the $\cos\theta-\phi$ map, only to facilitate the comparison with quantum MC results).


\vspace{5mm}
\begin{center}
{\bf III. Evolution of system structure with $N$}
\end{center}

Below we present a simple classical theory for the formation of the bilayer at $T=0$.
The theory predicts that, on increasing the number $N$ of particles, first a spherical layer forms where clusters are located at the vertices of an icosahedron;
subsequently, a second layer appears, concentric to the first one and at a larger distance from the center of the trap, where clusters sit at the vertices of a dodecahedron.
The theory is formulated for a lattice-gas model, but it is reasonable to expect that in a system of bosons the bilayer forms by the same energetic mechanism and according to the same filling scheme.

\begin{figure}[t]
\includegraphics[width=6cm,height=6cm]{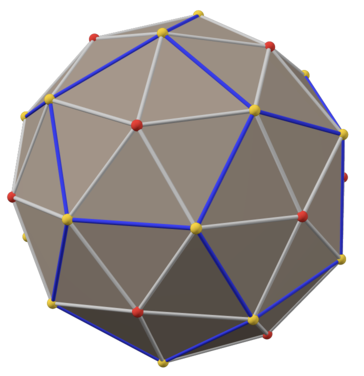}\qquad\qquad
\includegraphics[width=6cm,height=6cm]{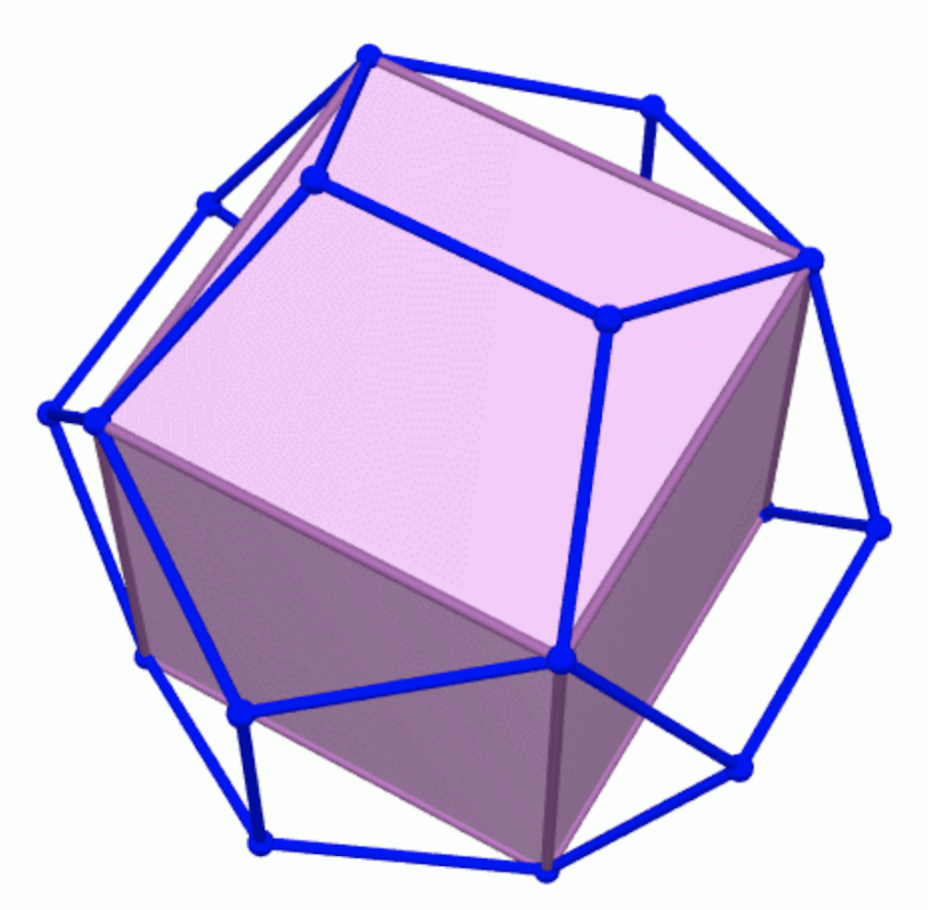}
\caption{Left: Pentakis dodecahedron (12 red vertices and 20 yellow vertices). Right: One of the five distinct cubes that can be formed from the vertices of a dodecahedron.}
\label{fig:model}
\end{figure}

A shell, or spherical layer, is modeled through a geodesic grid.
In particular, we consider the grid formed by the vertices and edges of a pentakis dodecahedron (PD), see Fig.~\ref{fig:model}a.
The sites of the grid are of two types:
Fivefold-coordinated, pentavalent sites (the red dots in the figure) and hexavalent sites (the yellow dots).
While the pentavalent sites (also referred to as A-type in the following) form a regular icosahedron, the hexavalent sites form a regular dodecahedron.
There are five distinct ways to draw a cube from the vertices of a dodecahedron (Fig.~\ref{fig:model}b):
For any such choice, the hexavalent sites can be divided in two groups:
cubic, or B-type sites (8 in total) and co-cubic, C-type sites (12).

Our theory applies for the classical model derived, as a limiting case, from the extended Bose-Hubbard model on the PD grid, studied in Ref.~\cite{prestipino2021}.
The Hamiltonian of the latter model is:
\begin{equation}
H=-t\sum_{\langle i,j\rangle}\left(a_i^\dagger a_j+a_j^\dagger a_i\right)+\frac{U}{2}\sum_in_i(n_i-1)+V\sum_{\langle i,j\rangle}n_in_j-\mu\sum_in_i\,,
\label{eq1sm}
\end{equation}
where $a_i$ and $a_i^\dagger$ are bosonic field operators and $n_i=a_i^\dagger a_i$ (with $i=1,\ldots,32$). $t>0$ is the hopping parameter, $U>0$ is the interaction energy between two particles sitting on the same site, $V>0$ is the interaction energy between two nearest-neighbor particles, and $\mu$ is the chemical potential.
By varying $\mu$, we can tune the number $N$ of particles in the system at equilibrium (with $N=\langle\sum_in_i\rangle$).
For $t=0$ the Bose-Hubbard model reduces to a classical lattice-gas model.
In the following, we examine precisely this limit, but releasing the constraint of single site occupancy analyzed in Ref.~\cite{prestipino2021}, since we wish to treat particles with a soft core.

Let us first consider the case of a single shell, corresponding to an external potential which is very sharp around the minimum at $R$.
To investigate the system structure it is convenient to assume, from a mean-field perspective, that sites of same type have the same occupation.
For fixed values of $T$ ($=0$) and $\mu$, the actual occupation will be such as to minimize the grand potential $\Omega=E-\mu N$, where $E$ is the total energy.
As $\mu$ increases for $U\rightarrow\infty$, we know from Ref.~\onlinecite{prestipino2021} that A-type sites are filled first ($N=12,0<\mu<3V$), then A and B sites ($N=20,3V<\mu<9V/2$), then A and C sites ($N=24,9V/2<\mu<6V$), and finally the whole grid ($N=32,\mu>6V$).
When $U$ is finite, each site can be occupied by multiple particles and the ``phases'' become many more.
The explicit calculation of $\Omega$ for the phases with simplest structure shows that, as far as $U>3V$, the sequence of phases at small $\mu$ remains identical to $U\rightarrow\infty$.
Conversely, for $U$ smaller than $3V$ there is a range of $\mu$ values where a few icosahedral cluster ``crystals'' occur in sequence, until eventually also the B and C sites start to be filled.
Therefore, when $U/V$ and $\mu$ are not too large, the system prefers to form icosahedral cluster crystals rather than having particles distributed on several ``sublattices''.

Let $n_\alpha=0,1,2,\ldots$ be the number of particles in every site of type $\alpha$ (with $\alpha=\,$A, B, C).
In order to catalog the system phases in ascending $\mu$ order, we must minimize (with respect to $n_{\rm A},n_{\rm B},n_{\rm C}$) the grand potential
\begin{eqnarray}
\Omega&=&\frac{U}{2}\left[12n_{\rm A}(n_{\rm A}-1)+8n_{\rm B}(n_{\rm B}-1)+12n_{\rm C}(n_{\rm C}-1)\right]
\nonumber \\
&+&\frac{V}{2}\left[12n_{\rm A}(2n_{\rm B}+3n_{\rm C})+8n_{\rm B}(3n_{\rm A}+3n_{\rm C})+12n_{\rm C}(3n_{\rm A}+2n_{\rm B}+n_{\rm C})\right]-\mu(12n_{\rm A}+8n_{\rm B}+12n_{\rm C})\,.
\label{eq2sm}
\end{eqnarray}
Minimization of \eqref{eq2sm} is cumbersome and is thus left to the computer.
We thus confirm that icosahedral cluster crystals ($n_{\rm A}>1,n_{\rm B}=n_{\rm C}=0$) are only present when $U<3V$.
The smaller $U$ relative to $3V$, the wider the $\mu$ interval where the population of A sites grows with $\mu$, while B and C sites remain empty.

Next we move to the case where the confining potential is no longer sharp.
Particles are now prompted to explore distances from the trap center ($O$) other than the potential minimum ($R$).
As suggested by the simulation, we imagine that particles are allowed to fill the sites of a second PD grid, concentric to the first grid and {\em external} to it, oriented such that the A sites of the second grid are lined up radially with those of the first grid.
We want that the filling of second-grid sites starts only after that an icosahedral crystal of $n$-clusters (with $n\approx 10$) has grown in the first grid.
Clearly, it is the external field that discourages the simultaneous filling of both grids:
while the first-grid sites are at the minimum of $u_{\rm ext}(r)$, the second-grid sites are more distant from $O$, so that a particle in the second grid pays an additional energy $\epsilon$ relative to a particle in the first grid.

Let $V'>0$ be the interaction between two nearest-neighbor sites of the second grid, $W>0$ the cost for having two ``twin'' sites (namely, two sites being lined up radially) simultaneously occupied, and $Z>0$ the interaction between an occupied first-grid site, say $i$, and an occupied site of the second grid that is neighbor of the twin site of $i$ (we assume that the radial separation between the grids is not too large, so that the separation between two nearest-neighbor sites in the second grid is just a little larger than the separation between two nearest-neighbor sites in the first grid).
Calling $n_\alpha'$ the occupations of second-grid sites, the grand potential of the bilayer is
\begin{eqnarray}
\Omega&=&\frac{U}{2}\left[12n_{\rm A}(n_{\rm A}-1)+8n_{\rm B}(n_{\rm B}-1)+12n_{\rm C}(n_{\rm C}-1)\right]
\nonumber \\
&+&\frac{V}{2}\left[12n_{\rm A}(2n_{\rm B}+3n_{\rm C})+8n_{\rm B}(3n_{\rm A}+3n_{\rm C})+12n_{\rm C}(3n_{\rm A}+2n_{\rm B}+n_{\rm C})\right]-\mu(12n_{\rm A}+8n_{\rm B}+12n_{\rm C})
\nonumber \\
&+&\frac{U}{2}\left[12n'_{\rm A}(n'_{\rm A}-1)+8n'_{\rm B}(n'_{\rm B}-1)+12n'_{\rm C}(n'_{\rm C}-1)\right]
\nonumber \\
&+&\frac{V'}{2}\left[12n'_{\rm A}(2n'_{\rm B}+3n'_{\rm C})+8n'_{\rm B}(3n'_{\rm A}+3n'_{\rm C})+12n'_{\rm C}(3n'_{\rm A}+2n'_{\rm B}+n'_{\rm C})\right]-(\mu-\epsilon)(12n'_{\rm A}+8n'_{\rm B}+12n'_{\rm C})
\nonumber \\
&+&W\left(12n_{\rm A}n'_{\rm A}+8n_{\rm B}n'_{\rm B}+12n_{\rm C}n'_{\rm C}\right)+Z\left[12n_{\rm A}(2n'_{\rm B}+3n'_{\rm C})+8n_{\rm B}(3n'_{\rm A}+3n'_{\rm C})+12n_{\rm C}(3n'_{\rm A}+2n'_{\rm B}+n'_{\rm C})\right]\,.
\nonumber \\
\label{eq3sm}
\end{eqnarray}
In order to reproduce the physics of the problem, we need to choose values for $V',W$, and $Z$ in a reasonable way.
To this purpose, we will be guided by the separation between the sites involved in the interaction, under the assumption that the strength of the interparticle repulsion is a decreasing function of distance.
Two twin sites are at relatively small separation from each other, hence we expect that $V<W\lesssim U$.
The distance between nearest-neighbor sites is a little larger in the second grid than in the first grid, hence $V'\lesssim V$.
Finally, the distance between two sites with $Z$ interaction is still larger, then we assume $0<Z<V'$.

We have minimized Eq.~(\ref{eq3sm}) with respect to $n_{\rm A},n_{\rm B},n_{\rm C},n_{\rm A}',n_{\rm B}',n_{\rm C}'$, choosing $U=2.5V,W=2V,V'=0.9V,Z=0.5V$, and $\epsilon=10V$.
By increasing $\mu/V$ progressively, we obtain the following results:
\begin{enumerate}
\item For $\mu<0$ both grids are empty ($N=0$);

\item For $0<\mu/V\lesssim 26.5$ we observe the gradual filling of the pentavalent sites of the first grid, until each one contains 11 particles ($N=132$);

\item Upon further increasing $\mu$, while the occupation of the A sites of the first grid keeps growing, the $B$ and $C$ sites of the second grid start to be filled:
first, B sites only ($n_{\rm A}'=0,n_{\rm B}'=1,n_{\rm C}'=0$), then C sites only ($n_{\rm A}'=0,n_{\rm B}'=0,n_{\rm C}'=1$), finally both of them ($n_{\rm A}'=0,n_{\rm B}'=1,n_{\rm C}'=1$).
At the end of this process, $\mu/V$ has grown to 35 and $n_{\rm A}$ to 13.

\item As soon as $\mu/V$ overcomes 35, a second particle appears in each B and C site of the second grid, according to the same scheme as before:
first in B sites only ($n_{\rm B}'=2,n_{\rm C}'=1$), then in C sites only ($n_{\rm B}'=1,n_{\rm C}'=2$), finally in both of them ($n_{\rm B}'=2,n_{\rm C}'=2$).
And so on.

\item When $\mu/V$ has eventually reached 154, we have $n_{\rm A}=48,n_{\rm B}'=14,n_{\rm C}'=14$.
All the other sites (i.e., the B and C sites of the first grid, and the A sites of the second grid) are empty.
Overall, the number of particles in the first grid is always larger than in the second grid.
\end{enumerate}
It is worth noting that, for $U$ substantially smaller than $2.5V$, things go differently from as described above.

Having established that in our simple model the bilayer is being filled according to the same scheme found in the simulation, we take $n_{\rm B}=n_{\rm C}=n_{\rm A}'=0$ in Eq.~(\ref{eq3sm}) and calculate the coordinates of the $\Omega$ minimum analytically, assuming {\em real} values for $n_{\rm A},n_{\rm B}',n_{\rm C}'$ (hereafter denoted as $x,y,z$).
The function to minimize is then:
\begin{eqnarray}
\Omega(x,y,z)&=&6Ux(x-1)-12\mu x+4Uy(y-1)+6Uz(z-1)+24V'yz+6V'z^2
\nonumber \\
&&-8(\mu-\epsilon)y-12(\mu-\epsilon)z+12Zx(2y+3z)\,.
\label{eq4sm}
\end{eqnarray}
Upon requiring the vanishing of first-order derivatives, we obtain the following set of linear equations:
\begin{equation}
\left\{
\begin{array}{l}
2Ux+4Zy+6Zz=U+2\mu \\
6Zx+2Uy+6V'z=U+2(\mu-\epsilon) \\
6Zx+4V'y+2(U+V')z=U+2(\mu-\epsilon)\,.
\end{array}
\right.
\label{eq5sm}
\end{equation}
Subtracting the third equation from the second, we obtain $y=z$ (providing $U\ne 2V'$), meaning that in this approximate treatment the filling of the B and C sites of the second layer is simultaneous.
Finally, replacing $y=z$ in the first and second equations, we get:
\begin{equation}
x=\frac{U+2\mu-10Zy}{2U}\,\,\,\,\,\,{\rm and}\,\,\,\,\,\,y=z=\frac{U^2+2(U-3Z)\mu-2U\epsilon-3UZ}{2(U^2+3UV'-15Z^2)}
\label{eq6sm}
\end{equation}
(providing $U^2+3UV'-15Z^2\ne 0$).
When $U^2+3UV'-15Z^2>0$ and $U-3Z>0$ (which is true for $U=2.5V,V'=0.9V$ and $Z=0.5V$), in order that $y>0$ we must require that
\begin{equation}
\mu>\frac{U\epsilon}{U-3Z}-\frac{U}{2}\equiv\mu_0\,.
\label{eq7sm}
\end{equation}
We note that $U>3Z$ and $Z<V'$ imply $U^2+3UV'-15Z^2>0$.
For the same parameters in the above example, we have $y>0$ for $\mu>23.8$ with $x=10$ at $\mu=23.8$.
These values are close to the true numbers (the small difference is due to $x,y,z$ being real rather than integer variables).

The particle numbers in the two layers are $N_1=12x$ and $N_2=20y$.
For $\mu\rightarrow\infty$, we get from Eq.~(\ref{eq6sm}) that
\begin{equation}
\frac{N_1}{N_2}\rightarrow\frac{3}{5}\frac{U+3V'-5Z}{U-3Z}\,.
\label{eq8sm}
\end{equation}
Obviously, the latter result is only valid so far as filling schemes different from that of the above example can be excluded.
Moreover, Eq.~(\ref{eq8sm}) does not account for the fact that, if $\mu$ becomes very large, a third layer will actually form, then followed by the onset of a fourth layer, and so on.

It remains to see under what further conditions, if any, the stationary point (\ref{eq6sm}) indeed corresponds to a minimum point.
To this aim, we need to show that the Hessian matrix $H(x,y,z)$ of (\ref{eq4sm}), as computed in the stationary point, is positive definite (a matrix is positive definite iff its eigenvalues are all positive, or iff the leading principal minors are all positive --- Sylvester criterion).
We immediately see that the Hessian matrix is actually a constant matrix,
\begin{equation}
H=
\begin{pmatrix}
12U & 24Z & 36Z\\
24Z & 8U & 24V'\\
36Z & 24V' & 12(U+V')
\end{pmatrix}
\,.
\label{eq9}
\end{equation}
Its leading principal minors are $12U,96(U^2-6Z^2)$, and $1152(U-2V')(U^2+3UV'-15Z^2)$.
The first minor is clearly positive.
The second minor is positive if $U>\sqrt{6}Z$, which is true for $U>3Z$.
Finally, the third minor (i.e., the determinant) is positive if $U>2V'$ and $U^2+3UV'-15Z^2>0$, which is true for $U>3Z$ and $V'>Z$.

We add a couple of final considerations.
\begin{enumerate}
\item We may ask why, for a not-too-sharp external potential, the second layer forms {\em externally} to the first layer, rather than inside it.
Inside the first layer there is little space, except possibly at $O$, to form a cluster at the right distance from the first layer;
instead, there is plenty of room outside and the interaction between particles located in different layers (not lined up radially) is on average weaker --- since their average separation is larger;

\item It follows from the previous numerical example that in order to have the desired scenario (i.e., icosahedral order in the first layer and dodecahedral order in the second layer), in addition to $U<3V$ we also need that $U>\max\{2V',3Z\}$.
Moreover, Eq.~(\ref{eq7sm}) indicates that $\epsilon$ must not be too small, since otherwise $\mu_0<0$.
In particular, since we need that $2V'<U<3V$, a flat interparticle repulsion cannot produce the wanted scenario within the present model.
This deficiency is a limitation of the lattice model, which is simply too rudimentary to produce, with a square-barrier potential, the same phenomenology seen in the simulation.
\end{enumerate}


\vspace{5mm}
\begin{center}
{\bf IV. Zonal superfluid fraction and superfluid density}
\end{center}

The definitions of superfluid density and superfluid fraction can be extended to have a spacial dependence when the system is separated into different regions, as detailed in \cite{ciardi2022b}. Here, we show how the derivation applies to the present case of a three-dimensional system.

In \cite{ciardi2022b}, the superfluid density is introduced for 2D systems as
\begin{equation}\label{eq:rhoslocal2D}
	\rho_s(\textbf{r}) = \frac{4 m^2}{\hbar^2 \beta}\frac{ \langle A A(\textbf{r}) \rangle - \langle A \rangle \langle A(\textbf{r}) \rangle }{r^2} \, ,
\end{equation}
where $A$ is the total area surrounded by worldlines in a given configuration, and
\begin{equation}\label{eq:azlocal2D}
	A(\textbf{r}) =  \frac{1}{2} \sum_{i=1}^{N} \sum_{m=0}^{M-1} \textbf{r} \times \textbf{r}_{i}^{m+1} \delta(\textbf{r}-\textbf{r}_{i}^{m})
\end{equation}
is the contribution to the area from a small region around point $\textbf{r}$, with $\textbf{r}_i^j$ the bead corresponding to particle $i$ on the $j$-th time slice. We use the notation $\textbf{v} \times \textbf{w} = v_x w_y - v_y w_x$ for 2D vectors.

When extending the definition to 3D systems, \eqref{eq:azlocal2D} is generalized to
\begin{equation}\label{eq:azlocal}
	\textbf{A}(\textbf{r}) =  \frac{1}{2} \sum_{i=1}^{N} \sum_{m=0}^{M-1} \textbf{r} \times \textbf{r}_{i}^{m+1} \delta(\textbf{r}-\textbf{r}_{i}^{m}) \,, 
\end{equation}
with each component of $\textbf{A}$ now measuring the area surrounded by worldlines projected onto the orthogonal plane. We must also be careful to distinguish the three-dimensional vector $\textbf{r}$ from the one appearing in the denominator of \eqref{eq:rhoslocal2D}, which represents the distance from the axis that we are considering to calculate the moment of inertia. For example, for the superfluid density around the $z$ axis, we must write
\begin{equation}\label{eq:rhoslocal}
	\rho_s^{(z)}(\textbf{r}) = \frac{4 m^2}{\hbar^2 \beta}\frac{ \langle A_z A_z(\textbf{r}) \rangle - \langle A_z \rangle \langle A_z(\textbf{r}) \rangle }{r_\perp^2} \, ,
\end{equation}
where we are now taking the $z$ component of the area, and $r_\perp = \sqrt{x^2 + y^2}$. We then have the definition of the superfluid fraction in region $k$ as 
\begin{equation}
f_{s,k}^{(z)} = \frac{1}{I_{cl,k}^{(z)}} \int_k d\textbf{r} \; \rho_s^{(z)}(\textbf{r}) r_\perp^2 \, ,
\end{equation}
where $I_{cl,k}$ is the classical moment of inertia, measured around the $z$ axis, of particles in region $k$.

In the simulation, we directly measure the integral $\int_k d\textbf{r} \; \rho_s^{(z)}(\textbf{r}) r_\perp^2$ by sampling the contributions of $\langle A_z A_z(\textbf{r}) \rangle - \langle A_z \rangle \langle A_z(\textbf{r}) \rangle$ to a given region. To associate a value of superfluid density to the entire region, we can introduce the constant $\rho_{s,k}^{(z)}$, defined to be such that
\begin{equation}
    \int_k d\textbf{r} \; \rho_s^{(z)}(\textbf{r}) r_\perp^2 = \int_k d\textbf{r} \; \rho_{s,k}^{(z)} r_\perp^2 \,.
\end{equation}
Replacing $r_\perp = r \sin\theta$, we are then left with
\begin{equation}
    f_{s,k}^{(z)} = \frac{\rho_{s,k}^{(z)}}{I_{cl,k}^{(z)}} \int_{r_k}^{R_k} dr \, r^4 \int d\Omega \sin^2\theta = \frac{\rho_{s,k}^{(z)}}{I_{cl,k}^{(z)}} \frac{8\pi}{3} \frac{R_k^5 - r_k^5}{5}
\end{equation}
(having called $r_k$ and $R_k$ the lesser and greater radius of the spherical shell, respectively), which we can invert to obtain $\rho_{s,k}^{(z)}$. For spherical shells with small thickness, this approximates $\rho_{s}^{(z)}(r)$.

Finally, we can introduce the integrated superfluid radial density
\begin{equation}
n_s^{(z)}(r) = r^2 \int{\rm d}\theta\,{\rm d}\phi \, \sin \theta \, \rho_{s}^{(z)}(r) \,,
\end{equation}
which directly compares to $n(r)$ introduced in the main text. We write $n_s(r)$ to generically refer to one or all the components of the integrated superfluid radial density.

\end{document}